\documentclass[pdflatex,iicol,sn-mathphys-ay]{sn-jnl}

\usepackage[T1]{fontenc}
\usepackage[utf8]{inputenc}
\usepackage{graphicx}
\graphicspath{{figures/}}
\usepackage{xurl}
\usepackage{multirow}
\usepackage{amsmath,amssymb,amsfonts}
\usepackage{booktabs}
\usepackage{makecell}
\usepackage{xcolor}
\usepackage{xspace}
\usepackage{listings}
\usepackage{algorithm}
\usepackage{algorithmicx}
\usepackage{algpseudocode}
\usepackage{siunitx}
\usepackage[title]{appendix}

\newcommand{\hilumi}{HL-LHC\xspace}
\newcommand{\cuda}{\textsc{CUDACPP}\xspace}
\newcommand{\helas}{\textsc{HELAS}\xspace}
\newcommand{\amcnlo}{\textsc{MadGraph5\_aMC@NLO}\xspace}
\newcommand{\mgfive}{MG5aMC\xspace}
\newcommand{\ttbar}{\ensuremath{\mathrm{t}\bar{\mathrm{t}}}\xspace}
\newcommand{\gluglu}{\ensuremath{\mathrm{gg}}\xspace}

\newcommand{\ggttg}{\ensuremath{\gluglu\rightarrow\ttbar\,\mathrm{g}}\xspace}
\newcommand{\ME}{$|{\cal M}|^2$\xspace}
\newcommand{\jamp}{\texttt{JAMP}\xspace}
\newcommand{\aie}{AI Engine\xspace}

\theoremstyle{thmstyleone}

\theoremstyle{thmstyletwo}

\theoremstyle{thmstylethree}

\begin{document}

\title[Cascade Pipeline for MC Matrix Elements on AMD Versal AI Engines]%
{Cascade Pipeline for Leading-Order Matrix Element Evaluation
on AMD Versal AI Engine Arrays}

\author*[1]{\fnm{P.}  \sur{Leguina}}\email{leguinapelayo@uniovi.es}
\author[3]{\fnm{C.}   \sur{Vico Villalba}}
\author[2]{\fnm{F.}   \sur{Herv\'{a}s \'{A}lvarez}}
\author[2]{\fnm{H.}   \sur{Guti\'{e}rrez Arance}}
\author[1]{\fnm{S.}   \sur{Folgueras}}
\author[1]{\fnm{J.}   \sur{Fern\'{a}ndez Men\'{e}ndez}}
\author[2]{\fnm{L.}   \sur{Fiorini}}
\author[2]{\fnm{A.}   \sur{Valero}}
\author[2]{\fnm{F.}   \sur{Carri\'{o}}}
\author[2]{\fnm{A.}   \sur{Oyanguren}}

\affil*[1]{\orgname{Universidad de Oviedo -- ICTEA},
           \orgaddress{\city{Oviedo}, \country{Spain}}}
\affil[2]{\orgname{Instituto de F\'{i}sica Corpuscular (IFIC)},
          \orgdiv{Universitat de Val\`{e}ncia -- CSIC},
          \orgaddress{\city{Valencia}, \country{Spain}}}
\affil[3]{\orgname{Rice University},
          \orgaddress{\city{Houston}, \state{Texas}, \country{USA}}}

\abstract{%
A major computational bottleneck in modern High Energy Physics event generators arises from the integration of the matrix element, which requires repeated evaluations at different phase-space points to cover all possible initial- and final-state configurations. As the Large Hadron Collider enters its High-Luminosity phase, the demand for energy-efficient acceleration is expected to exceed the limits of conventional CPU scaling, motivating the use of highly parallel computing platforms such as graphics processing units (GPUs). In this work, we present an alternative approach based on a cascade pipeline architecture for evaluating leading-order matrix elements of the \ggttg process on AMD Versal AI Engine (\aie) arrays. Due to the 16\,kB per-tile program memory constraint, the computation is decomposed into a five-stage pipeline, with stages communicating via a wavefunction-token protocol over the on-chip cascade interface. Mapping 80 independent pipelines onto the 400 \aie tiles of the VCK190 platform yields a projected throughput of $1.0\times10^6$ matrix element evaluations per second at 54.8\,W, corresponding to a $34\times$ speedup over a single CPU core and a $7.7\times$ improvement in energy efficiency. Numerical agreement with the \amcnlo double-precision reference is validated at the parts-per-million level in mean relative error.
}


\maketitle


\section{Introduction}\label{sec:intro}

Physics event generators can be used to model proton--proton collisions at the Large Hadron
Collider (LHC) by integrating quantum transition probabilities over
a given phase space using Monte Carlo (MC)
techniques~\citep{HSFPhysicsEventGeneratorWG:2020gxw}.  These generators encode
standard model (SM) interactions in computational form and produce quantitative
predictions for any specified process.  The calculations are naturally parallel: the
same squared matrix element (\ME) is evaluated independently at each sampled
phase-space point.

As the LHC enters its High-Luminosity phase (\hilumi), computing demand grows
beyond linear scaling trends, while available computing budgets remain approximately
flat~\citep{HSFPhysicsEventGeneratorWG:2020gxw}. The matrix element evaluation
step can account for 30--40\% of the total event generation CPU time, particularly for processes
that include additional real emissions in the matrix element calculation. We refer to such processes
as multi-jet processes. These multi-jet processes constitute the primary computational
bottleneck~\citep{Valassi:2021ljk}. Closing the resulting computational gap requires
hardware acceleration beyond conventional CPU improvements.

Several approaches address this bottleneck through parallelism.  The \cuda
plugin for \amcnlo~\citep{Valassi:2021ljk,Valassi:2025xfn} exploits GPU and vectorized-CPU
architectures, achieving $100\times$--$1000\times$ speedup over
single-threaded CPU on complex processes~\citep{Hageboeck:2023blb,Wettersten:2025hrb}. 
Field-Programmable Gate Arrays (FPGAs) offer an alternative approach based on custom fixed-function datapaths and lower power per-tile operation. Initial studies in~\citet{Barbone:2023mul} demonstrated the feasibility of FPGA-based acceleration for Monte Carlo simulation on Xilinx Alveo and AMD Versal platforms. More recently,~\citet{Gutierrez:2025particles} reported the port of the $e^+e^-\to\mu^+\mu^-$ matrix element from \mgfive to FPGA using high-level synthesis.

Our work builds on the automated implementation of matrix element computations within the \amcnlo (\mgfive) framework. For the specific derivation of the functions used to compute the matrix elements, we refer the reader to the documentation about the \helas framework in \citep{Murayama:1992gi}, which is used to perform the amplitude calculations for a given set of initial and final state, as well as its implementation within the \amcnlo framework in \citep{Alwall:2014hca}.

For the physics use case, we focus on the process in which two gluons fuse to produce a quark--antiquark top pair (\ttbar). To demonstrate the feasibility of handling more complex processes, we additionally include one extra emission in the matrix element calculation. We refer to this full process as \ggttg throughout the paper. The functions that integrate the matrix element and are ported into the FPGA are obtained at the lowest order possible in \mgfive. To our knowledge, no prior work has mapped such a process onto the 
AMD Versal \aie array.  


This paper presents several contributions. First, a memory-driven five-stage cascade pipeline 
is introduced, partitioning the full \ggttg matrix element computation across five \aie tiles connected 
via a wavefunction-token protocol over the 384-bit cascade interface. 
Second, the pipeline stages adhere to a deterministic cascade architecture, characterized by identical loop structures, unconditional writes, and statically matched token counts, thereby ensuring deadlock-free operation by construction. Third, the complete \helas amplitude library is ported to \aie vector intrinsics, incorporating complex-division reduction, vectorized kernels, and binary-indexed helicity caching. Finally, an 80-pipeline deployment across the 400~\aie tiles of the VCK190 platform is projected to achieve $1.0\times10^6$ matrix element evaluations per second at 54.8\,W \aie-domain power. Physics equivalence is validated against the \mgfive double-precision reference over 1\,000 phase-space points.


The remainder of this paper is organised as follows.
Section~\ref{sec:related} reviews Monte Carlo event generation, GPU/CPU
acceleration efforts, and prior FPGA work.
In Section~\ref{sec:architecture}, we present the cascade pipeline
architecture, token design, cascade contract, and HELAS adaptations.
Section~\ref{sec:experimental} specifies the experimental setup.
The results are reported in Section~\ref{sec:results}, including program
memory, precision, throughput, power, and resource utilisation.
Section~\ref{sec:comparison} provides a cross-platform performance comparison.
Section~\ref{sec:conclusion} concludes and outlines future work.


\section{Related Work}\label{sec:related}

This section reviews existing work in three areas: Monte Carlo event
generation frameworks, GPU and vectorised-CPU acceleration, and FPGA-based
approaches.

\subsection{Monte Carlo event generation and the HELAS library}

Modern event generators such as \mgfive automate the
calculation of scattering amplitudes from Feynman rules. The \helas
library provides a standardized set of subroutines that 
allow to represent each Feynman diagram as a
sequence of wavefunction construction and vertex contraction calls.  The
\ME is obtained by summing diagram amplitudes into
colour-flow amplitude vectors (\jamp) and contracting with a colour matrix.

For the \ggttg process at leading order, \mgfive generates 16~Feynman
diagrams, 6~colour-flow amplitudes, and 32~helicity configurations.  The
complete \helas call graph uses 10~distinct functions spanning four categories:
external wavefunction generators (incoming fermion, outgoing fermion, vector
boson), fermion--fermion--vector vertices, triple-gluon vertices, and
four-gluon contact vertices.  This function set and the 5-particle final state
make \ggttg a compact yet non-trivial benchmark for hardware acceleration.

\subsection{GPU and vectorized-CPU acceleration}

The \cuda plugin for \mgfive implements automatic code generation for NVIDIA GPUs and AVX2/AVX-512 vector
CPUs.  On an NVIDIA A100 GPU, the plugin achieves $2.18\times10^7$~matrix
elements per second for \ggttg in single-precision mode. Vectorized CPU implementations provide up to $16\times$ 
speedup on a single core relative to scalar code.  \citet{Hageboeck:2023blb} reported results from
the first alpha release, and \citet{Wettersten:2025hrb,Wettersten:2023ekm} extended GPU acceleration
beyond leading order to next-to-leading-order calculations.

A TensorFlow-based framework for GPU-accelerated matrix element computation, presented by \citet{Carrazza:2021gpx} and known as MadFlow, establishes the GPU throughput and baseline capabilities across several architectures, which are subsequently compared.

\subsection{FPGA and Versal acceleration}

\citet{Barbone:2023mul} demonstrated FPGA acceleration for Monte Carlo
simulation on both a Xilinx Alveo card and the AMD Versal platform.
For the specific $e^+e^-\to\mu^+\mu^-$ scattering process,
\citet{Gutierrez:2025particles} reported a \mgfive-to-FPGA port based on
high-level synthesis.  These studies established the feasibility of the
approach on simpler kernels and lower-complexity processes.

Two aspects distinguish the present work from the above. First, the target
process (\ggttg) involves 16~diagrams and 10~HELAS functions, a substantially
more complex computation than the lower-diagram-count
$e^+e^-\to\mu^+\mu^-$ process.  Second, the 16\,kB program memory
constraint per \aie tile becomes the primary architectural driver for
multi-diagram processes and requires a systematic multi-tile partitioning
strategy; prior Versal work used single-tile solutions that were sufficient
for simpler processes.


\section{Architecture and Implementation}\label{sec:architecture}

\begin{figure*}[t]
\centering
\includegraphics[width=\textwidth]{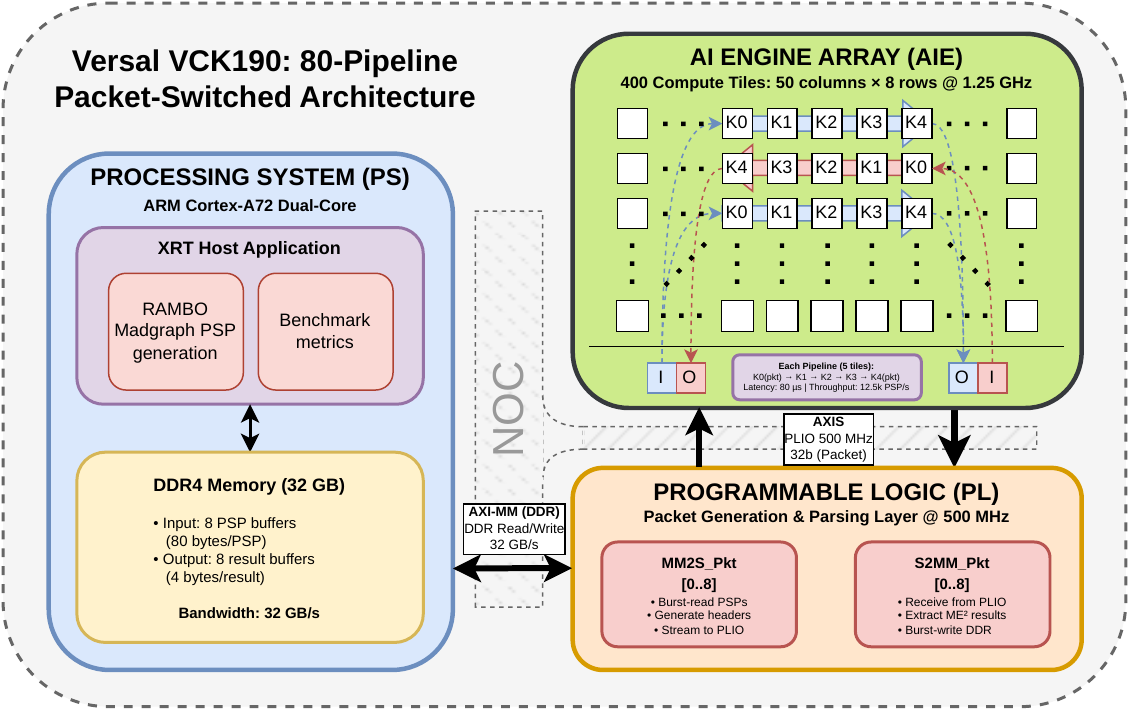}
\caption{Full system-on-chip architecture on the VCK190.
The Processing System (PS, ARM Cortex-A72) hosts the runtime and
phase-space generator; DDR4 provides input/output memory.
The Programmable Logic (PL) implements 10 memory-to-stream and
10 stream-to-memory HLS data movers at 500\,MHz, each distributing
phase-space points to a group of eight pipelines via packet-switched
stream routing.
The \aie array contains 80~five-tile cascade pipelines in
10~column groups of 8~rows.  Each PLIO port is placed at the centre
column of its group, limiting every pipeline entry to at most two
stream-switch hops.}
\label{fig:full_arch}
\end{figure*}

Figure~\ref{fig:full_arch} shows the complete system-on-chip deployment
on the VCK190 platform.  The following subsections describe the target
platform, the program-memory constraint that motivates the design, the
five-stage cascade pipeline, the inter-tile communication protocol, and the
HELAS library adaptation.

\subsection{Target platform: AMD Versal XCVC1902}\label{sec:platform}

The AMD Versal AI Core Series VCK190 evaluation kit contains the XCVC1902
Adaptive Compute Acceleration Platform
(ACAP)~\citep{AMD:VersalAICore,AMD:AIEArchReference}.
The device integrates three compute domains on a single die: a dual-core ARM
Cortex-A72 application processor (Processing System, PS) for system control
and host orchestration; a programmable logic (PL) fabric with 899\,840 LUTs,
1\,968 DSP58 engines, and 158\,MB of embedded memory (UltraRAM + BRAM); and
an \aie array of 400~tiles arranged in a $50\times8$ grid, clocked at
1.25\,GHz~\citep{AMD:VersalAICore,AMD:AIEArchReference}.

\subsection{AI Engine tile microarchitecture}\label{sec:aie_tile}

Each \aie tile contains a 512-bit SIMD vector unit supporting
single-precision (float32) and fixed-point arithmetic, a 32-bit scalar RISC
core for control flow, 32\,kB of local data memory (with 128\,kB addressable
via neighbour tiles), and 16\,kB of dedicated program memory (PM).  The
program-memory limit is a hard architectural constraint: a kernel that
exceeds 16\,kB cannot be compiled.  Unlike GPUs or CPUs, where code size is
effectively unconstrained, this limit determines what computation can reside on
a single tile.

\subsection{Cascade interface}\label{sec:cascade}

Adjacent tiles in the same row are connected by a 384-bit unidirectional
accumulator bus, the cascade interface.  This interface provides deterministic,
zero-overhead transfer of one 384-bit word per clock cycle, with no
flow-control signals.  A 4-entry hardware FIFO decouples producer and consumer
by up to four cycles, absorbing minor timing skew.  The cascade direction is
fixed per row: even rows (0, 2, 4, 6) cascade left-to-right; odd rows
(1, 3, 5, 7) cascade right-to-left.  The resulting bandwidth of
$384\,\text{bits/cycle}\times 1.25\,\text{GHz} = 60\,\text{GB/s}$ far
exceeds the stream I/O bandwidth from the shim.

\subsection{Program-memory constraint analysis}\label{sec:pm_analysis}

The monolithic kernel that implements all 16~Feynman diagrams of the \ggttg
process requires a compiler-reported program-memory footprint of
approximately 38\,kB---a factor of 2.4 above the
16\,kB tile limit.  This excess arises because the 10~\helas vertex functions,
their supporting logic, and the token serialization code cannot coexist in a
single tile.

The partitioning is further constrained by the fact that certain pairs of
\helas functions, when combined with the necessary kernel logic overhead, fill
or exceed the 16\,kB limit.  Each kernel therefore includes only the subset of
\helas headers required by its assigned diagrams; co-locating incompatible
function pairs in one kernel exceeds the compiler-reported PM budget.

A second program memory optimization involves \emph{deferred evaluation}: the
off-shell boson generator \texttt{FFV1P0\_3} (2\,680~bytes) was initially placed
in the first pipeline stage, but profiling showed that this configuration
required 17.8\,kB.  Since that function is consumed only by the fourth stage,
moving it there while retaining the Stage~1 triple-gluon current generator
\texttt{VVV1P0\_1}
reduces the first stage from 17.8\,kB to 15.5\,kB ($-13\%$) and simultaneously
shrinks the inter-tile token by one wavefunction, saving two cascade beats per
helicity iteration.

\subsection{Five-stage cascade pipeline}\label{sec:pipeline}

\begin{figure*}[t]
\centering
\includegraphics[width=\textwidth]{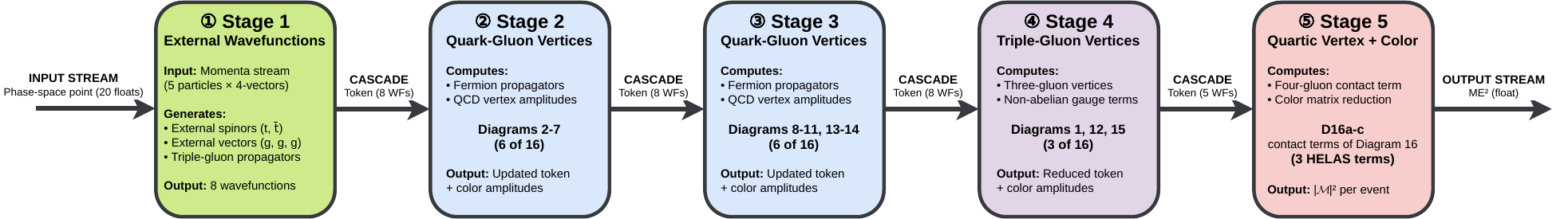}
\caption{Single five-stage cascade pipeline occupying one row of five \aie
tiles.  Stage~1 receives a phase-space point, generates eight external
wavefunctions (five physical, three precomputed triple-gluon currents), and
emits a wavefunction token via the cascade interface.
Stages~2 and~3 evaluate the fermion--vector diagram groups (D2--D7 and
D8--D11, D13--D14), accumulating partial colour-flow amplitudes.
Stage~4 evaluates the triple-gluon diagrams D1, D12, D15, including
the deferred \texttt{FFV1P0\_3} vertex.
Stage~5 evaluates the four-gluon contact contributions D16a--c of diagram
D16, contracts the colour matrix, and emits the squared modulus of the matrix
element.
The cascade token carries eight wavefunctions through Stages~1--4 and
is reduced to five between Stages~4 and~5.}
\label{fig:single_pipe}
\end{figure*}

The \ggttg computation is decomposed into five sequential pipeline stages,
each mapped to one \aie tile in a single row.  Figure~\ref{fig:single_pipe}
illustrates the data flow through a single pipeline instance.

Only the boundary stages (first and last) connect to stream I/O; all interior
stages communicate exclusively via the cascade interface.  This separation
makes the pipeline design independent of the external I/O topology.

Table~\ref{tab:diag_assign} specifies the Feynman diagram and \helas function
assignment for each stage.

\begin{table*}[t]
\caption{Diagram-to-stage assignment for the five-stage \ggttg pipeline.
Stage~5 includes the three D16 sub-diagrams arising from four-gluon contact
vertices.}
\label{tab:diag_assign}
\centering
\small
\setlength{\tabcolsep}{5pt}
\begin{tabular}{@{}clp{6.2cm}l@{}}
\toprule
Stage & Diagrams & HELAS functions & Role \\
\midrule
1 & ---               & \texttt{vxxxxx}, \texttt{oxxxxx}, \texttt{ixxxxx}, \texttt{VVV1P0\_1}            & Wavefunction gen.\ \& token init.\ \\[4pt]
2 & D2--D7            & \texttt{FFV1\_1}, \texttt{FFV1\_2}, \texttt{FFV1\_0}                              & Fermion--vector eval., partial \jamp \\[4pt]
3 & D8--D11, D13--D14 & \texttt{FFV1\_1}, \texttt{FFV1\_2}, \texttt{FFV1\_0}                              & Fermion--vector eval., partial \jamp \\[4pt]
4 & D1, D12, D15      & \texttt{FFV1P0\_3}, \texttt{VVV1\_0}                                              & Deferred boson, triple-gluon amp.\ \\[4pt]
5 & D16a--c           & \texttt{VVVV1P0\_1}, \texttt{VVVV3P0\_1}, \texttt{VVVV4P0\_1}                    & Four-gluon contact \& colour red.\ \\
\bottomrule
\end{tabular}
\end{table*}

The twelve fermion--vector diagrams are split evenly between Stages~2 and~3 to
keep each below 80\% PM utilization, providing headroom for compiler
optimization.  The off-shell object retained in Stage~1 is therefore the
triple-gluon current \texttt{VVV1P0\_1}, whereas the deferred function moved to
Stage~4 is the off-shell boson generator \texttt{FFV1P0\_3}.

\subsection{Cascade token protocol}\label{sec:token}

All inter-stage state is passed via cascade tokens---structured
sequences of cascade beats encoding wavefunctions and partial colour-flow
amplitudes.  Two token formats are used.

The extended token, employed between Stages~1 and~4, carries five external
wavefunctions ($w_0$--$w_4$), three precomputed triple-gluon propagators, and
the six-element colour-flow amplitude vector:
\begin{equation}
  \underbrace{8 \times 2}_{\text{wf.}} + \underbrace{2}_{\text{amp.}}
  = 18\ \text{cascade beats/helicity}.
  \label{eq:extended_token}
\end{equation}
Each beat transfers four complex floats (32\,bytes) via the 384-bit
accumulator.  Wavefunction components occupy lanes 0--5; lanes~6--7 are
zero-padded for alignment with the 8-wide vector unit.
Stage~4 is the final recipient of this extended token format; it consumes the
three precomputed triple-gluon propagators to evaluate the triple-gluon vertex
diagrams (D1, D12, D15) before emitting the reduced token to Stage~5.

The reduced token, used between Stages~4 and~5, carries only the five
external wavefunctions plus amplitudes:
\begin{equation}
  5 \times 2 + 2 = 12\ \text{cascade beats per helicity},
  \label{eq:standard_token}
\end{equation}
since the off-shell boson vertex is evaluated locally by Stage~4, as described
in Section~\ref{sec:pm_analysis}.

Stages~2 and~3 each independently recompute six propagators from the external
wavefunctions in the token rather than forwarding them.  Forwarding these
propagators would add 12~cascade beats per helicity, increasing the token size
by 67\%.  Recomputation costs approximately 200~cycles per propagator but
keeps the token at 18~beats, reducing cascade bandwidth usage by approximately
40\%.  This is a deliberate computation-versus-communication trade-off that
keeps the throughput model compute-dominated rather than cascade-limited.

\subsection{Deterministic cascade contract}\label{sec:contract}

To ensure deadlock-free operation by construction without flow-control
hardware, all stages in a pipeline obey three rules:
\begin{enumerate}
  \item \textbf{Identical loop structure.}  Every stage iterates over all
    phase-space points in the batch and all 32~helicity configurations in fixed
    order.  No early exits or data-dependent iteration counts are permitted.
  \item \textbf{Unconditional cascade writes.}  Each iteration writes exactly
    one token.  Conditional writes would cause FIFO drift between adjacent
    tiles, eventually producing deadlock.
  \item \textbf{Statically matched counts.}  The number of tokens written by
    each producer equals the number of tokens read by its consumer.
\end{enumerate}
These rules are enforced by a token-level API: dedicated write and read
functions for each token format serve as the sole inter-stage communication
interface.  The 4-entry cascade FIFO absorbs minor cycle-to-cycle skew from
branch mispredictions in the scalar unit, provided the loop structure is
preserved.

\subsection{HELAS adaptation for AIE vector intrinsics}\label{sec:helas_adapt}

The \helas amplitude library, originally generated by \mgfive for scalar
double-precision C++, is ported to the \aie vector instruction set.  Five
key adaptations are described below.

The reference \mgfive code represents each wavefunction as a 6-element
array of double-precision complex numbers.  The \aie port maps these into
8-wide single-precision complex vectors (the native width of the \aie SIMD
unit).  Wavefunction physics components occupy lanes 0--5.  Lanes~6--7 store
precomputed coupling constants ($iV_3$ and $iV_4$), multiplied once per
wavefunction construction.  Downstream fermion--vector vertex functions read
these cached values directly, eliminating one complex multiplication per call.
When wavefunctions traverse the cascade interface, lanes~6--7 are
zero-padded because the cascade does not support partial writes; the receiving
stage recomputes the cache as needed.

Propagator denominators of the form $(p^2 - M^2 + iM\Gamma)^{-1}$ require
complex inversion.  The reference code uses Smith's method, which involves two
real divisions.  The \aie implementation replaces this with a single
real-valued reciprocal:
\begin{equation}
  \frac{1}{a + ib} = \frac{a}{a^2+b^2} - i\frac{b}{a^2+b^2},
\end{equation}
computing $1/(a^2+b^2)$ with one hardware reciprocal instruction, followed by
two multiplies.  This reduces the division count per Breit--Wigner propagator
from two to one.

The dominant computation pattern in the fermion--vector vertex functions
(off-shell fermion production) is a $4\times4$ complex linear system on spinor
components.  This is refactored into a shared core function that computes four
outputs via fused multiply-accumulate chains:
\begin{equation}
  F_i = \sum_{j} C_{ij}\, S_j, \quad i \in \{0,1,2,3\},
\end{equation}
where $C_{ij}$ are coupling-dependent coefficients.  A precomputation structure
evaluates 12~linear combinations of the gauge-boson momentum components once
per vertex call using scalar arithmetic, eliminating redundant
recomputation across the four spinor outputs.

Recomputing all five external wavefunctions for each of 32~helicity
configurations would require 160~evaluations per phase-space point.  Since
each particle takes helicity $\pm1$, the implementation precomputes
10~wavefunctions ($2\times5$) and selects them via bit-indexed lookup:
\begin{align}
  \mathrm{bit}_k &= (h \gg k) \mathbin{\&} 1, \notag \\
  w_k &= \mathrm{bit}_k \mathbin{?}\ w_k^{+}\ :\ w_k^{-},
  \quad k \in \{0,\ldots,4\}.
\end{align}
This reduces evaluations from 160 to 10 per phase-space point---a $16\times$
reduction---at the cost of 10~vector registers.

Finally, the colour reduction computes
$|M|^2 = \sum_{ij} C_{ij}\,\mathrm{Re}(\jamp_i^*\,\jamp_j)$
using the $6\times6$ colour matrix for \ggttg.  All divisions by the
colour normalisation factor are folded into compile-time constants (6~diagonal
and 15~off-diagonal terms), reducing the inner loop to 21~fused multiply-adds
with zero runtime divisions.  Spin averaging ($\times 1/256$) is a single
scalar multiply on the final sum.

\subsection{Scalable 80-pipeline array}\label{sec:array}

The 400~\aie tiles are organized into $10\,\text{column groups}\times
8\,\text{rows} = 80$ independent five-tile pipelines.  Each column group
occupies five consecutive columns; the five pipeline stages fill exactly those
columns.

Tile placement respects the fixed cascade direction of each row: even rows
cascade left-to-right (Stage~1 at the leftmost column) and odd rows cascade
right-to-left (Stage~1 at the rightmost column).  The kernel bodies are
identical across all 80~pipelines; only the placement constraints differ.

Each column group is served by one PLIO input/output pair.  The PLIO port is
placed at the centre column of its group (at shim columns 6, 10, 14, \ldots,
42---all valid sites on XCVC1902), so that every pipeline entry point within
the group is at most two hops away through the \aie stream-switch
infrastructure.  A packet-split node fans the incoming stream to all eight
entry stages in the group; a packet-merge node collects the eight output
values.  Packet ID bits encode the row index (0--7), routing each phase-space
point to the correct pipeline.  This topology minimizes routing congestion
because inbound and outbound streams traverse minimal paths within each column
group.

The complete deployment spans three layers as shown in
Figure~\ref{fig:full_arch}: the Processing System (ARM A72) handles
phase-space generation and runtime orchestration; the Programmable Logic
implements 10~memory-to-stream and 10~stream-to-memory HLS data movers at
500\,MHz, generating packet-header AXI-Stream transfers at initiation
interval~1; and the \aie array hosts all 80~five-tile cascade pipelines,
fully utilizing all 400~tiles.


\section{Experimental Setup}\label{sec:experimental}

This section specifies the hardware platforms, software tools, and benchmark
configuration used for all measurements reported in
Sections~\ref{sec:results}--\ref{sec:comparison}.

\subsection{AIE target platform}

The primary target is the AMD Versal VCK190 evaluation board containing the
XCVC1902 ACAP~\citep{AMD:VersalAICore,AMD:AIEArchReference}.  The \aie array comprises 400~tiles at 1.25\,GHz.  The PL
fabric runs at 500\,MHz for HLS data movers.  DDR4 provides 32\,GB
of device memory accessed via the Network-on-Chip (NoC) AXI-MM interface.
The Vivado power report attributes 54.8\,W to the \aie domain (79\% of the
68.9\,W dynamic power) and 82.7\,W to the full chip (static + dynamic),
as measured by the on-chip power manager.

\subsection{Software and tools}

The \aie kernels are compiled with AMD Vitis 2024.1 using default optimization
flags.  PL synthesis and place-and-route of HLS data movers use Vivado 2024.1.
The host runtime is Xilinx Runtime (XRT) on Linux, running on the ARM
Cortex-A72.  Functional validation uses the \aie x86 simulator; latency
estimates are obtained from the cycle-approximate \aie simulator.  The
reference generator is \amcnlo v3.5.6~\citep{Alwall:2014hca} with the \cuda
plugin v1.00.00~\citep{Valassi:2025xfn} for GPU and vectorised-CPU baselines.

\subsection{Benchmark process and dataset}

The target process is \ggttg at leading order, generated by \mgfive.  The
centre-of-mass energy is 1.5\,TeV.  Phase-space points are produced by a RAMBO
uniform phase-space generator running on the ARM A72 processor.  Each point
encodes the four-momenta of five particles~$(g_1, g_2, t, \bar{t}, g_3)$ as
20~single-precision floats (80~bytes).

Precision validation uses 1\,000~phase-space points comparing the float32
\aie output against the float64 \mgfive reference.  Throughput measurements
use batches of 655\,360~phase-space points to ensure steady-state behavior.
Single-pipeline latency is reported by the cycle-approximate \aie simulator,
the standard schedule-level timing estimator in the Vitis \aie workflow.
The projected full-array figure is consequently a deployment model, not a
directly measured end-to-end hardware throughput.

The 80-pipeline projection assumes linear scaling: the 80~pipelines are
architecturally independent, with separate cascade chains and packet-switched
I/O paths.  The complete 80-pipeline design passed the \aie compilation flow
and Vivado implementation, yielding the reported post-implementation resource
and power figures.  The packet-switched I/O topology is therefore present in
the implemented design, but hardware end-to-end timing of the full array was
not collected; all throughput figures derive from the cycle-approximate
simulator.

\subsection{Reference platforms for comparison}\label{sec:ref_platforms}

Two reference platforms are used.  The CPU baseline is an Intel Core i5-10600
at 3.30\,GHz, running the \mgfive benchmark binary in single-threaded mode
with aggressive optimization (\texttt{-Ofast}) in double precision.  This
figure serves as a contextual single-core CPU reference rather than a tuned
vectorised-CPU result.  Power is measured via Intel RAPL (package domain).

The GPU reference is an NVIDIA A100-PCIE-40GB (250\,W TDP), with CUDA 13.0
and driver 580.95.05.  The \cuda benchmark binary is executed in
single-precision mode.  Power is measured via \texttt{nvidia-smi} (average
draw during kernel execution).

For a fair comparison, only the matrix element computation time is used for
throughput calculation on all platforms, excluding random number generation and
phase-space sampling overhead.


\section{Results}\label{sec:results}

This section presents the program-memory, programmable-logic, precision,
throughput, latency, and power results.

\subsection{Program memory utilisation}\label{sec:res_pm}

\begin{figure}[ht]
\centering
\includegraphics[width=\columnwidth]{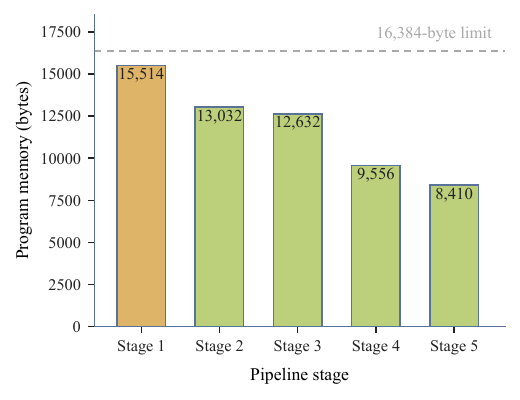}
\caption{Program-memory usage per pipeline stage as reported by the \aie
compiler.  The dashed line marks the 16\,384-byte per-tile hardware limit.
Stage~1 is the most constrained stage, reaching 15\,514~bytes (94.7\% of the
available program memory).}
\label{fig:pm_results}
\end{figure}

Figure~\ref{fig:pm_results} shows the program memory reported by the \aie
compiler for each pipeline stage after applying all optimisations described in
Section~\ref{sec:architecture}.  All five stages fit within the 16\,kB limit.
Stage~1 is the most constrained at 94.7\% utilisation.  The deferred-evaluation
optimisation that moved the off-shell boson vertex from Stage~1 to Stage~4
reduced the first stage from 17.8\,kB to 15.5\,kB, a saving of 2.3\,kB
($-13\%$).

\subsection{Programmable logic resource utilization}\label{sec:res_pl}

Table~\ref{tab:pl_util} summarizes the PL resource consumption for the
complete design, including 10~memory-to-stream and 10~stream-to-memory HLS
data movers plus platform infrastructure.  PL utilization is modest: LUT
usage is 4.72\%, register usage is 2.87\%, and BRAM usage is 16.0\%.  The
design leaves the majority of PL resources available for additional logic if
needed.

\begin{table}[ht]
\caption{Programmable-logic resource utilization for the full VCK190 design.}
\label{tab:pl_util}
\begin{tabular}{@{}lrrl@{}}
\toprule
Resource & Used & Available & Utilisation (\%) \\
\midrule
CLB LUTs         & 42\,499   & 899\,840   & 4.72 \\
Registers        & 51\,675   & 1\,799\,680 & 2.87 \\
Block RAM tiles  & 155       & 967        & 16.03 \\
Slices           & 11\,150   & 112\,480   & 9.91 \\
Bonded I/Os      & 382       & 692        & 55.20 \\
\bottomrule
\end{tabular}
\end{table}

\subsection{Numerical precision}\label{sec:res_prec}

Physics equivalence is validated by comparing the float32 \aie output against
the float64 \mgfive reference at a centre-of-mass energy of 1.5\,TeV.
Validation covers four levels:
external wavefunctions ($w_0$--$w_4$), diagram-level amplitudes (D1--D16),
the full colour-flow amplitude vector, and the final \ME output.
Table~\ref{tab:precision} summarizes the precision statistics.

\begin{table}[ht]
\caption{Precision of the \aie float32 implementation relative to the
\mgfive float64 reference over 1\,000~phase-space points.}
\label{tab:precision}
\begin{tabular}{@{}lr@{}}
\toprule
Metric & Value \\
\midrule
Mean absolute error          & $1.50 \times 10^{-7}$ \\
Maximum absolute error       & $9.30 \times 10^{-5}$ \\
Mean relative error          & $1.43 \times 10^{-6}$ (1.4\,ppm) \\
Maximum relative error       & $1.68 \times 10^{-4}$ (168\,ppm) \\
\bottomrule
\end{tabular}
\end{table}

The mean relative error of 1.4\,ppm is consistent with the theoretical
precision of IEEE~754 single-precision arithmetic ($2^{-23} \approx
1.19\times10^{-7}$ machine epsilon) applied to a multi-step calculation
involving 16~Feynman diagrams with hundreds of floating-point operations each.
The worst-case relative error of 168\,ppm occurs at a phase-space point with
a relatively large matrix element value (0.554), consistent with accumulated
rounding in the colour reduction.

The float32 relative precision of $\sim10^{-6}$ is several orders of magnitude
below the physical uncertainties in leading-order calculations ($\sim1\%$ from
scale variation and parton distribution functions).  Systematic rounding
errors, rather than random noise, arise from the deterministic float32
accumulation order.  This margin supports the use of the float32
implementation for leading-order event generation, where physics systematics
dominate the numerical error budget.  All 1\,000~phase-space points produce
identical final \ME outputs across repeated runs, confirming output-level
determinism of the cascade pipeline.

\subsection{Throughput and latency}\label{sec:res_throughput}

The cycle-approximate \aie simulator reports a single-pipeline latency of
80\,$\mu$s from phase-space point injection at the input stream to \ME
read-back at the output stream, corresponding to a single-pipeline throughput
of $1.25\times10^4$~ME/s.

With 80~architecturally independent pipelines under the modeled I/O load
(separate cascade chains, separate packet-switched I/O, no shared memory or
synchronization bottlenecks),
the projected full-array throughput under linear scaling is:
\begin{equation}
  \text{Throughput}_{\text{80-pipe}}
  = \frac{80}{80\,\mu\text{s}}
  = 1.0 \times 10^6\ \text{ME/s}.
  \label{eq:throughput}
\end{equation}

This full-array value is therefore a projected deployment throughput rather
than a directly measured end-to-end hardware throughput.

The throughput model indicates that the architecture is \emph{compute-dominated}
under the measured single-pipeline latency and calculated I/O load.  Each
pipeline consumes 80~bytes (one phase-space point) and emits 4~bytes (one \ME
result) plus 8~bytes of packet overhead.  Per-trunk stream I/O capacity at
500\,MHz and 32-bit width is 2\,GB/s.  With eight pipelines sharing one trunk:
\begin{align}
  &\text{I/O util.\ (in)} \notag \\
  &\quad= \frac{8 \times 80\,\text{B} \times
    1.25 \times 10^4\,\text{s}^{-1}}{2\,\text{GB/s}} \approx 0.4\%, \notag \\
  &\text{I/O util.\ (out)} \ll 0.1\%.
\end{align}
Phase-space point injection time per packet (21~beats at 500\,MHz) is
42\,ns, compared to 80\,$\mu$s of compute---a ratio of approximately
$1\!:\!1\,900$.  These ratios indicate that further optimization should focus
primarily on kernel compute latency.

\subsection{Power and energy}\label{sec:res_power}

The Vivado power report attributes 54.8\,W to the \aie domain (79\% of
dynamic power), 65.5\,W to the full accelerator subsystem
(AIE + NoC/DDR + I/O), and 82.7\,W to the total chip (including static).
The energy per matrix element, using the AIE-domain figure, is:
\begin{equation}
  E_{\text{AIE}} = \frac{54.8\,\text{W}}{1.0 \times 10^6\,\text{ME/s}}
  = 54.8\,\mu\text{J/ME}.
\end{equation}
Using the full-chip power gives 82.7\,$\mu$J/ME, providing an upper bound
that includes all static and infrastructure power.


\section{Comparison}\label{sec:comparison}

This section compares the \aie cascade pipeline against CPU and GPU platforms
at three levels: throughput, power, and energy per matrix element.

\subsection{Platform overview}

Table~\ref{tab:comparison} summarizes the comparison data used in this work.
CPU and GPU values are measured, whereas the 80-pipeline \aie throughput is a
projection obtained from the measured single-pipeline latency under a
linear-scaling assumption.  All throughput values correspond to the pure
matrix element computation time, excluding phase-space sampling and random
number generation.  Unless otherwise noted, the headline \aie comparison uses
\aie-domain power because it isolates the accelerator array responsible for
matrix element evaluation; the full-chip value is retained as a separate
system-level upper bound.  The GPU power entry corresponds to the average draw
during kernel execution.  Throughout this section, throughput is reported in
matrix elements per second (ME/s), \ME denotes the squared modulus of the matrix element, and
E/ME denotes energy per matrix element in $\mu$J/ME.

\begin{table*}[t]
\caption{Cross-platform comparison for \ggttg matrix element evaluation.
CPU and GPU values are measured using the \mgfive \cuda
benchmark~\citep{Valassi:2025xfn}.  \aie power refers to the AI Engine
domain (54.8\,W); the full VCK190 chip power is 82.7\,W.  The 80-pipeline
throughput is projected from the single-pipeline latency
(Section~\ref{sec:res_throughput}).  The single-pipeline power entry is an
effective per-pipeline share, not a direct measurement.  GPU power is the
average draw during kernel execution.}
\label{tab:comparison}
\centering
\small
\begin{tabular}{@{}lcccc@{}}
\toprule
Platform & Precision & Throughput (ME/s) & Power (W) & E/ME ($\mu$J) \\
\midrule
VCK190, 80 AIE pipelines            & fp32 & $1.0\times10^{6}$ & 54.8 & 54.8 \\
VCK190, 1 AIE pipeline (eff.\ share) & fp32 & $1.25\times10^{4}$ & 0.685 & 54.8 \\
i5-10600 (1 core, 3.3\,GHz)           & fp64 & $2.92\times10^{4}$ & 12.3 & 422  \\
A100-PCIE-40GB                        & fp32 & $2.18\times10^{7}$ & 159  & 7.3  \\
\bottomrule
\end{tabular}
\end{table*}

The effective-share row is included only to show that the linear-scaling model
preserves energy per matrix element when the measured \aie-domain power is
distributed across 80 identical pipelines.

\subsection{Throughput comparison}

The projected 80-pipeline \aie deployment reaches $1.0\times10^6$~ME/s, a
$34\times$ throughput improvement over the single-core CPU baseline
($2.92\times10^4$~ME/s).  This speedup is achieved with a $4.5\times$ power
increase (54.8\,W versus 12.3\,W).

The single-core CPU figure serves as a contextual reference from the
same benchmark workflow rather than a directly like-for-like fp32 or
vectorised-CPU comparison.

The NVIDIA A100 GPU achieves $2.18\times10^7$~ME/s, exceeding the \aie array by
a factor of $21.8$ in raw throughput.  This advantage reflects the much larger
compute budget of the A100 (6\,912~CUDA cores at up to 1.41\,GHz and
312~TFLOPS peak float32) compared with the 400~\aie tiles.

\subsection{Energy comparison}

Energy per matrix element provides a power-normalized comparison that accounts
for the different thermal envelopes of the platforms.  Relative to the
single-core CPU baseline, the \aie array consumes 54.8\,$\mu$J/ME versus
422\,$\mu$J/ME, corresponding to a $7.7\times$ improvement in energy
efficiency.  Relative to the GPU, the A100 reaches 7.3\,$\mu$J/ME and is thus
$7.5\times$ more energy-efficient per matrix element than the \aie, but it also
operates at $2.9\times$ higher absolute power (159\,W versus 54.8\,W).  The
GPU power entry corresponds to an average draw of 159\,W during kernel
execution, compared with 54.8\,W for the \aie domain and 12.3\,W for the CPU
baseline.  The \aie result therefore remains attractive in power-constrained
settings when peak GPU throughput is not required.

\subsection{Discussion}

Beyond throughput and energy, the main differences
between the platforms lie in software maturity, numerical precision,
available array capacity, and power envelope.

The GPU ecosystem (cudacpp, cuBLAS, compiler toolchains) has been optimized for
over 15~years, with the cudacpp plugin representing a mature, production-ready
implementation.  The \aie implementation is a first-generation prototype.
Latency reductions from helicity filtering (Section~\ref{sec:conclusion}),
deeper pipelining, or higher AIE clock rates on next-generation Versal devices
could narrow the throughput gap.

The CPU baseline uses double-precision arithmetic; the \aie and GPU use
single-precision.  The throughput comparison against the CPU should
therefore be read as contextual rather than strictly like-for-like.  The
measured ppm-level mean relative error of the \aie implementation is
consistent with IEEE~754 float32 and is adequate for leading-order
calculations.  Next-to-leading-order calculations may require mixed-precision
strategies.

All 400 \aie compute tiles of the XCVC1902 array are occupied by the 80
five-tile pipelines.  Further scaling requires either a larger \aie array (e.g., the Versal
Premium series with up to 472~\aie tiles) or a multi-device deployment.  The GPU scales to
multi-GPU configurations via standard MPI or NCCL parallelism.

The \aie-domain power of 54.8\,W operates within an 82.7\,W total chip
envelope.  This moderate chip-level power budget makes the design potentially
relevant to power-constrained accelerator deployments, including trigger or
online-reconstruction contexts at the LHC.  The A100 has a 250\,W thermal
design power and requires active cooling infrastructure.


\section{Conclusion}\label{sec:conclusion}

This work presents a cascade pipeline architecture for leading-order Monte
Carlo matrix element evaluation on AMD Versal \aie arrays.  For \ggttg, the
monolithic kernel footprint exceeds the 16\,kB per-tile program-memory budget,
motivating a five-stage decomposition in which the most constrained stage
(wavefunction generation) occupies 15\,514~bytes (94.7\%).  Wavefunctions,
precomputed propagators, and partial colour-flow amplitudes are transferred
between stages over the 384-bit cascade interface via a contract-based
deterministic protocol that ensures deadlock-free operation.  All 400~\aie
compute tiles of the XCVC1902 array are occupied by 80~independent five-tile
pipelines.

The projected array-level throughput---$34\times$ faster than a single CPU
core at 54.8\,$\mu$J/ME versus 422\,$\mu$J/ME---demonstrates that the
cascade architecture achieves meaningful energy efficiency gains while
preserving ppm-level agreement with the \mgfive double-precision reference.

The work has several limitations.  Only the leading-order
\ggttg process is demonstrated.  The reported single-pipeline latency is
simulator-derived rather than directly timed on hardware, and the extension of
the methodology to other processes is argued from the partitioning strategy
rather than demonstrated experimentally.  The float32 precision is adequate for
leading-order calculations but may require mixed-precision strategies at
next-to-leading order.  The NVIDIA A100 GPU outperforms the \aie array by
$21.8\times$ in raw throughput and $7.5\times$ in energy per matrix element,
reflecting the maturity of the GPU ecosystem and the larger compute budget of
the A100.

Several directions for future work follow from this design.  The most
direct improvement is helicity filtering: precomputing a good-helicity
bitmask at runtime reduces the 32-helicity inner loop to approximately
16~significant configurations, which is expected to provide roughly a
$2\times$ speedup without structural changes to the cascade contract.  The
partitioning methodology is expected to generalise to higher-multiplicity
processes---in particular $\ttbar ggg$, the multi-jet topologies most directly
relevant to LHC top-quark measurements---by extending the pipeline depth.  NLO
integration appears feasible since real-emission and virtual-loop corrections
employ the same HELAS function set, with the primary challenge being the
increased token complexity arising from loop-integral wavefunction products.
Finally, next-generation Versal devices with larger \aie arrays or higher
clock rates would increase the available pipeline count and throughput
proportionally.

\section*{Acknowledgements}
This work was supported by the Ministerio de Ciencia e Innovaci\'{o}n of Spain
and by the Generalitat Valenciana.

\section*{Statements and Declarations}

\subsection*{Competing interests}
The authors declare no competing interests, financial or non-financial, related
to this work.

\subsection*{Funding}
This work was supported by the Ministerio de Ciencia e Innovaci\'{o}n of Spain
and by the Generalitat Valenciana.

\subsection*{Author contributions}
P.~Leguina L\'{o}pez conceived and carried out the work, including the
hardware and software implementation, validation, analysis, visualisation, and
manuscript drafting.
J.~Fern\'{a}ndez Men\'{e}ndez contributed expertise in Monte Carlo event
generation and reviewed the manuscript.
C.~Vico Villalba contributed expertise in Monte Carlo event generation and
reviewed the manuscript.
S.~Folgueras contributed to project administration, supported the work through
funding acquisition, and reviewed the manuscript.
F.~Carri\'{o} contributed expertise on the AMD Versal hardware platform and
reviewed the manuscript.
A.~Valero and L.~Fiorini contributed through technical discussion, project
administration, and manuscript review.
H.~Guti\'{e}rrez Arance contributed through investigation, technical
discussion, and manuscript review.
F.~Herv\'{a}s \'{A}lvarez carried out a preliminary \aie baseline study.
A.~Oyanguren contributed to project administration and funding support.

\subsection*{Data availability}
The \LaTeX\ sources, figure files, and supporting analysis scripts are
available at the project repository.\footnote{\url{https://github.com/pleguina/Wavefunction-Token-Cascade-for-MC-Matrix-Elements-on-AMD-Versal-AI-Engines}}
Additional data are available from the corresponding author upon reasonable
request.

\bibliography{references}

\end{document}